# Manifold Learning of Four-dimensional Scanning Transmission Electron Microscopy

Xin Li[1,2], Ondrej E. Dyck[1,2], Mark P. Oxley[1,2], Andrew R. Lupini[1,2],

Leland McInnes[3], John Healy[3], Stephen Jesse[1,2], Sergei V. Kalinin[1,2]

[1]Center for Nanophase Materials Sciences, Oak Ridge National Laboratory, Oak Ridge TN 37831, USA

[2]Institute for Functional Imaging of Materials, Oak Ridge National Laboratory, Oak Ridge TN 37831, USA

[3]Tutte Institute for Mathematics and Computing, Ottawa, Canada

This manuscript has been authored by UT-Battelle, LLC under Contract No. DE-AC05- 00OR22725 with the U.S. Department of Energy. The United States Government retains and the publisher, by accepting the article for publication, acknowledges that the United States Government retains a non-exclusive, paid-up, irrevocable, world-wide license to publish or reproduce the published form of this manuscript, or allow others to do so, for United States Government purposes. The Department of Energy will provide public access to these results of federally sponsored research in accordance with the DOE Public Access Plan (http://energy.gov/downloads/doe-public-access-plan).

# Abstract

Four-dimensional scanning transmission electron microscopy (4D-STEM) of local atomic diffraction patterns is emerging as a powerful technique for probing intricate details of atomic structure and atomic electric fields. However, efficient processing and interpretation of large volumes of data remain challenging, especially for two-dimensional or light materials because the diffraction signal recorded on the pixelated arrays is weak. Here we employ data-driven manifold leaning approaches for straightforward visualization and exploration analysis of 4D-STEM datasets, distilling real-space neighboring effects on atomically resolved deflection patterns from single-layer graphene, with single dopant atoms, as recorded on a pixelated detector. These extracted patterns relate to both individual atom sites and sublattice structures, effectively discriminating single dopant anomalies via multi-mode views. We believe manifold learning analysis will accelerate physics discoveries coupled between data-rich imaging mechanisms and materials such as ferroelectric, topological spin and van der Waals heterostructures.

# Introduction

Over the past decade, Scanning Transmission Electron Microscopy (STEM) has seen a growing trend towards full capture of the information stream including amongst many others, three-dimensional (3D) electron tomography[1], four-dimensional (4D) ptychography[2], 4D phase plate STEM[3] and five-dimensional (5D) in-situ beam electron diffraction[4]. Efficient phase contrast imaging in STEM using a pixelated detector provides comprehensive information at each scanning location via the Ronchigram images[5] (In the rest of paper, we use the notation of Ronchigram to indicate the zero order disk from a beam diffraction pattern recorded on a pixelated detector). A large body of theoretical and experimental work has been conducted, revealing that 4D Ronchigram datasets enable, for example, super resolution[6,7] and three-dimensional imaging[8–10]. Aside from efficient phase imaging, Pennycook et al.[5] concluded that Ronchigram images should have greater sensitivity to electromagnetic fields than differential phase contrast imaging (DPC) which is currently considered as the state-of-the-art to directly visualize local electromagnetic fields[11–18]. With recent progress in high-sensitive area detectors[19], DPC has revealed atomic electric fields in crystal by both segmented detectors[20,21] and pixelated detectors with a simplified quantum theoretical interpretation[22].

Despite the above progress, a vast fraction of the information contained in the Ronchigram is currently left unutilized, due to the lack of understanding of how physical parameters such as instrumentation settings and material structures affect the deflection patterns (inhomogeneous electron intensity distribution recorded on the pixelated detector), necessitating a data-driven analytical framework without excessive prior domain knowledge. Here we investigate the deflection patterns and the associated real-space distributions over atomic structures via low-dimensional manifold learning of the Ronchgiram datasets. It is often the case that the underlying structure of the dataset as a whole can be described in terms of a much smaller number of latent features than the high dimensional set of measurements. The low-dimensional physical parameter space is translated onto a high dimensional response space by means of the imaging mechanisms. Aside from possible discontinuities due to material structure transitions and non-linear imaging transfer functions, points in close proximity in the physical parameter space will generally be in close proximity in the response space, forming a complex non-linear manifold. The non-linearity often renders linear dimension reduction methods[23] of limited usefulness. Manifold learning is a set of well-established non-linear techniques in the machine learning community for finding such latent structure in high dimensional data.

Denote the collected Ronchigram datasets over a scanned area as $\mathbf{X} = \{\mathbf{X_1}, \mathbf{X_2}, \ldots, \mathbf{X_n}\} \subseteq R^p$, where $n$ is the total number of scanned pixel locations and $p$ is the image size of an individual Ronchigram. We use the recently developed uniform manifold approximation and projection (UMAP[24]) method to represent the high-dimensional Ronchigram datasets via fuzzy topological sets with regard to the Riemannian metric. For straightforward visualization and exploration analysis, here UMAP optimizes the low-dimensional manifold representation $\mathbf{Y} = \{\mathbf{Y_1}, \mathbf{Y_2}, \ldots, \mathbf{Y_n}\} \subseteq R^d$, $d = 2,3$ of the original Ronchigram datasets $\mathbf{X}$, by minimizing the cross entropy of the two fuzzy set representations for **X** and **Y**. Efficient implementation of UMAP follows the work of LargeVis[25], by firstly constructing approximate nearest neighbor graph via random projection tree[26] and neighbor exploring[27], then solving the low-dimensional manifold embedding via probabilistic edge sampling and negative sampling[28]. Further analysis of the intrinsic structures within the manifold can be performed via the machine learning technique known as clustering. Trying to further separate manifold clusters and present them in a clearer way, Li et al. empirically proposed Graph-Bootstrapping[29] that iteratively reconstructs the nearest neighbor graph based on previous manifold positions and then recalculates manifold coordinates based on the reconstructed graph.

Remarkably, although the proposed manifold learning framework is purely data-driven without any prior bias regarding the material structure and instrumental modality, it reveals the real-space neighboring effects on deflection patterns in the Ronchigram sampled over single-layer graphene with single dopant atoms. These deflection patterns relate to both individual atomic site and the sublattice structure and can be used to effectively discriminate dopant anomalies via multiple-mode views. For the experimental datasets of size (64*64, 180*180), the computation took less than 10 minutes on a single workstation (Intel Xenon E5-1650V3, 32GB DDR3 RAM) to get the findings analyzed in this paper.

Figure 1 illustrates the flowchart of manifold-learning accelerated discovery of physics coupled between imaging mechanisms and material systems. High-dimensional Ronchigram datasets are projected into a low-dimensional manifold space for efficiently revealing and hierarchically representing rich features. With extracted patterns from manifold learning, deeper study can be conducted via adaptive experiment design. In addition, we believe this type of analysis could also stimulate the development of advanced machine learning algorithms. The driving goal of unsupervised learning is to discover unknown but interesting patterns within data. Due to the inherent absence of labels within the field of unsupervised machine learning, the ultimate validation of our work is the discovery of interesting, useful and externally validated results. External validation, such as on data with physical meaning, is an example of how we hope to develop mathematically sound and broadly applicable techniques, while simultaneously confirming the physical interpretations.

## Results

In this paper we focus on two specific example datasets, one synthetic dataset where the ground-truth input (the sample atomic positions and scattering factors) are known exactly, and an experimental dataset with several unknown parameters. The simulated data is a (37, 64, 120, 140) synthetic dataset over 37 x 64 probe positions with the individual Ronchigram of size of 120 X 140 pixels. The experiment consists of 64 x 64 probe positions with 180 x 180 Ronchigram pixels giving a (64, 64, 180, 180) experimental dataset.

## Manifold Clustering

In this work, we only consider the Euclidean distance as the distance metric. UMAP requires the tuning parameters of local neighborhood size and the effective minimum distance between embedded points, which we set to be 50 and 0 respectively for all manifold learning cases including the Graph-Bootstrapping procedure. We leave all the other tuning parameters of UMAP as default.

For the clean synthetic dataset, UMAP projects the graphene Ronchigram dataset into the shape of a hexagon in Figure 2a. With a clear manifold layout, based on visual inspection on manifold shape, we can simply perform spectral clustering[30] with 7 clusters. Supplementary Figure 1 contains manifold layouts from 25 reruns of UMAP on the synthetic dataset, where we overlay the same set of cluster labels in Figure 2a.

Figure 2b displays the UMAP manifold derived from the experimental dataset. We more often could not get a good guess on the number of clusters based on manifold shape due to noise and distortions associated in the experimental data (imagine Figure 2b without color labels). To further extract the cluster structure, we calculate bootstrapped UMAP manifold by first reconstructing the graph from UMAP manifold (Figure 2b) then recalculating the manifold embedding from the reconstructed graph as shown in Figure 2c. Since the number of clusters is related to the structure-pattern relationships that is an unknown parameter we would like to estimate, clustering on manifold should be generally based on local structure such as nearest neighbor and density. Here we utilize the hierarchical density estimate methods (HDBSCAN[31,32]) to perform clustering on the bootstrapped UMAP manifold. Mathematically, HDBSCAN relies on the mutual reachability distance: $D_{mreach,i}(a,b) = \max\{core_i(a), core_i(b), d(a,b)\}$, where $d(a,b)$ is the original metric distance between points $a$ and $b$, $core_i(x)$ is the core distance of a point $x$ to cover its $i$th nearest neighbor.

The primary tuning parameter of HDBSCAN is the minimum cluster size $k$. We follow a similar procedure in Ref[29] to choose $k$. We first consider all integer $k$ values in the range [20,149]. In supplementary Figure 2, we then plot the trend of total number of estimated clusters against every $k$ value and fit this trend by the exponential decay. We choose the $k$ in the tail region where the total number of clusters tends to be stable. Specifically, we set $k$ =86 in this case. We leave all the other tuning parameters of HDBSCAN as default.

HDBSCAN clustering is performed on bootstrapped UMAP manifold. Figure 2b,c show the clustering patterns over UMAP and bootstrapped UMAP manifolds with the same set of HDBSCAN cluster labels. We note that points with label of "-1" are the "outliers" identified by HDBSCAN that do not belong to any cluster. Supplementary Figure 3 contain manifold layouts from 25 reruns of UMAP on the experimental dataset, where we overlay the same set of cluster labels in Figure 2b,c .

## Synthetic Dataset Analysis

Figure 3a is the high-angle annular dark filed (HAADF) image of the synthetic dataset. Figure 3b displays the known, ground-truth atom positions overlaid on the spatial mapping of cluster labels derived from the manifold space in Figure 2a. There are 5 Si dopants with one 4-fold coordinated dopant located in the middle and the others located at the four corners. Most obviously, we can see clusters 0,2,5 and clusters 3,4,6 consist of two groups that are located around atom sites on the two sublattices of graphene respectively, while cluster 1 is mostly located in the space between atoms. Figure 3c displays the mean and standard deviation (Std) of Ronchigrams for each cluster with positions to the atom sites. We note that standard deviations ($10^{-6}$-$10^{-5}$) are one magnitude smaller than mean values ($10^{-4}$), showing the clustering accuracy. Arrows in Figure 3c are used to indicate that the electron beam is deflected towards the direction of atom nuclei which is consistent with previous work on unveiling atomic electric fields with DPC[21,22]. From Figure 3b,c, we further note that Ronchigrams with opposite deflection patterns are located at opposite sides of the two mirrored atom sites over the two sublattices, instead of the two sides of the same atom, revealing the real-space neighboring effects on deflection patterns. It is also worthwhile to point out the irregular real-space position arrangements of cluster labels located at Si dopants.

To check the above finding quantitatively, one can calculate the similarity loadings by calculating pairwise distances between the mean Ronchigram of the cluster and every individual Ronchigram. Figure 4 shows the similarity loadings (inverse of pairwise Euclidean distances) of clusters. For every cluster, the distribution of bright blobs in the similarity loading is consistent with cluster label positions. Comparing similarity loadings of clusters with opposite deflection patterns in Figure 4 (clusters 0 and 3, clusters 2 and 4, clusters 5 and 6), we see none of those bright blobs are located at opposite sides of the same atom site. Instead, they are located at the opposite sides of the two mirrored atom sites over sublattices. Effectively, the method decomposes deflection patterns into 2 classes that are spatially related to the sublattices, instead of being radially symmetric around each atom site. We conjecture that both individual atom site and its local neighbor atoms will affect the deflection pattern. Furthermore, we notice that all similarity loadings display a black patch at the 4-fold Si dopant position in the middle, indicating the deflection patterns around the Si dopant are significantly different from those around the carbon atom.

## Experimental Dataset Analysis

Figure 5a shows the HAADF image of graphene with a single 4-fold silicon dopant, overlaid with estimated atom positions and Figure 5b is an example of an as-acquired experimental Ronchigram where the deflection pattern is weak due to the monolayer structure of graphene. Nonetheless manifold learning could still distinguish intrinsic patterns in the real dataset without any signal pre-processing. Figure 5c displays the spatial distribution of cluster labels showing the sublattice structures.

Figure 6 shows the similarity loadings of the clusters. Based on the relative positions between atom sites and bright blobs as illustrated by the circle markers, we categorize clusters into sublattice A (clusters 0,1,5,6) and sublattice B (clusters 2,3,7,8). Again, all the similarity loadings display a black patch at the position of the 4-fold Silicon dopant, implying differences between Ronchigrams sampled around the Si dopant and those around the carbon atom. Supplementary Figure 4 shows the similarity loadings of clusters (clusters 4,9) whose spatial locations are not indicative to either sublattice.

Finally, trying to see the relative patterns in experimental Ronchigrams, we subtracted the mean of all Ronchigrams from the mean Ronchigram of each cluster in Figure 7. Supplementary 5 displays the original mean Ronchigrams of clusters with a manually tuned range of colorbar. By referring to the same cluster number, we can verify that deflection patterns in cluster Ronchigrams (Figure 7, Supplementary Figure 5) are consistent with associated cluster label positions to the atom sites as indicated in Figure 5c, Figure 6 and Supplementary Figure 4. Supplementary Figure 6 provides a compact illustration of deflection patterns in Ronchigrams and associated positions to the atom sites.

## Discussion

Over the past 10 years, several groups have made significant contributions towards visualization of electromagnetic fields via segmented and pixelated detectors. However, at the present time, a vast fraction of the information contained in the Ronchigram is still not fully exploited and there remains a lack in knowledge of how physical parameters affect the electron distribution on pixelated detectors. Here we utilize the data-driven manifold learning methodology, to spatially map the local inhomogeneities captured in the Ronchigram datasets and to do so without requiring a priori model or placing any constraints on the data. This machine learning method could provide a fast screening tool for domain experts to digest the Ronchigram patterns at large scale and build knowledge libraries of various effects made on Ronchigram. To illustrate the above point, Figure 8 shows the Ronchigram patterns from the synthetic dataset with -10 nm defocus. Otherwise, the defocused dataset was calculated as described in the methodology section. Under this defocus setting, the manifolds in Figure 8a,b only consist of two groups. The spatial distributions of cluster labels are no longer indicative of the graphene sublattice structure. To study the patterns of Ronchigram, attention should be paid on design of experiment, with careful control and recording of the experimental conditions.

We have shown that manifold learning is a highly efficient method to process and visualize large scale 4D-STEM datasets of atomically-resolved diffraction patterns. As we highlight in both the synthetic and experimental data analysis, even though diffraction pattern is very weak due to the monolayer structure of graphene, we can distill hidden information from Ronchigrams. Instead of the radial symmetry centered at each atom site, manifold learning analysis separates the Ronchigrams based on local symmetry dependent on the sublattices, indicating the real-space neighboring effects on deflection patterns. We propose that manifold learning and other related techniques will be extremely promising routes for analyzing the wealth of multidimensional data emerging from a new generation of electron microscopes, to denoise, interpolate, and explore the data that will be of interest to many fields such as ferroelectric, topological spin and van der Waals heterostructures.

In particular, this approach may open a pathway for low dose imaging and effective algorithms for atomic manipulations[33–37] that currently rely on sequential imaging and probe positioning. Methods based on the Ronchigram will have the potential for more efficient use of the signal for thin samples than the readily interpretable HAADF image[5], effectively allowing us to use a single point signal as a proxy for the local structure and obviate the need for sub-scan or Fourier transform based feedback[38]. Efficient manifold learning method might be adapted to provide a fast measure of whether a certain area matches a known cluster or represents an anomaly, with potential benefits for real-time feedback and control.

## Methods

Simulation of Ronchigrams was carried out using the quantum excitation of phonons option in the $\mu$STEM package[39]. Calculations were carried out on a 768 x768 pixel mesh with the unit cell tiled to form a supercell approximately 42 x 37 Å. A total 80 Monte Carlo passes were used. A probe forming aperture of 30 mrad was assumed.

Experimental 4D imaging was acquired using a Nion UltraSTEM 100 microscope operated at the 60 kV accelerating voltage and the 30 mrad convergence angle. The pixel dwell time was 20 ms.

CVD-grown graphene samples were transferred from Cu foil to TEM grids and cleaned via a wet transfer method and baking in an $ArO_2$ environment as described elsewhere[40]. The dopant atom was inserted *in situ* as described elsewhere[41].

The spectral clustering algorithm has been included in the Python sklearn package at http://scikit-learn.org/stable. The HDBSCAN Python package can be found at https://github.com/scikit-learn-contrib/hdbscan. The UMAP Python package can be found at https://github.com/lmcinnes/umap

## Data availability

The data that supports the findings of this study are available at https://doi.org/10.6084/m9.figshare.7416317. Python scripts and the GUI based on Python Bokeh Library are available at https://github.com/nonmin/4D-STEM.

## Acknowledgements

This research was partially supported (XL, OED, SJ,SVK) at the Center for Nanophase Materials Sciences, which is a US DOE Office of Science User Facility. Part of this work (MPO, ARL) was supported by the Office of Basic Energy Sciences, Materials Sciences and Engineering Division, US Department of Energy. LM and JH acknowledge support from Tutte Institute for Mathematics and Computing, Canada. We gratefully acknowledge Myron D. Kapetanakis for providing the structure files used in the simulation.

## Competing Interests

## Author Contributions

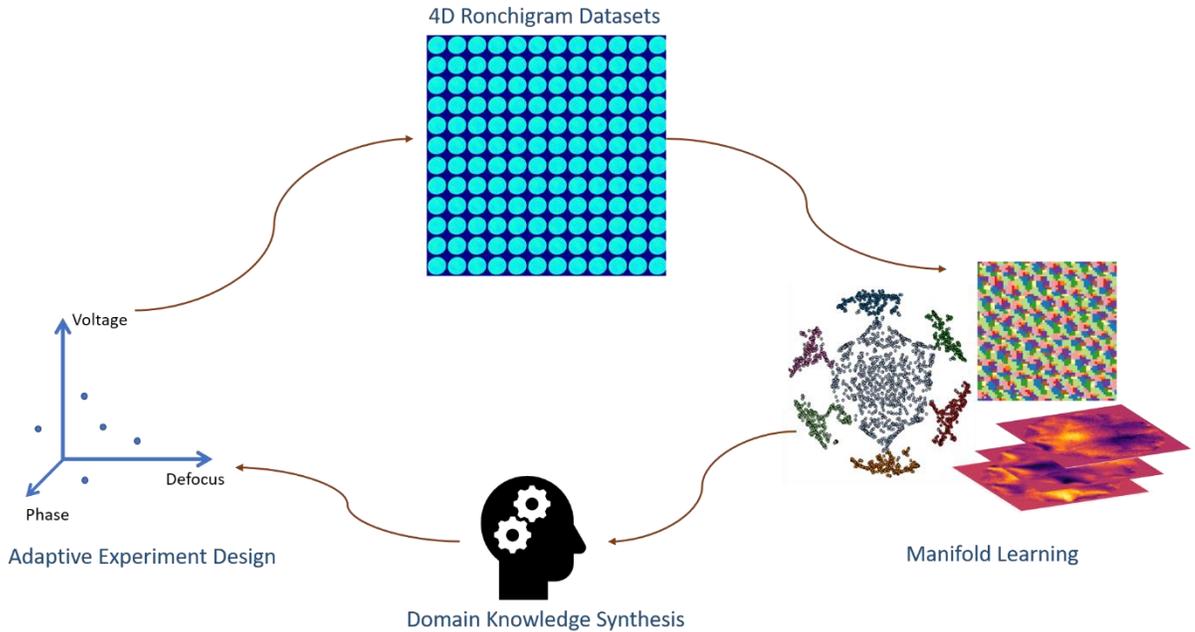

**Figure 1.** Schematic of manifold learning of Ronchigram datasets. The low-dimensional physical parameter space (such as defocus, accelerating voltage, material structure phases) is translated onto a high dimensional Ronchigram response space by the imaging mechanisms of microscope. High-dimensional and large-scale Ronchigram datasets are projected into a low-dimensional manifold space for efficiently revealing and hierarchically representing rich features. With extracted patterns from manifold learning, deeper study can be conducted via adaptive experiment design.

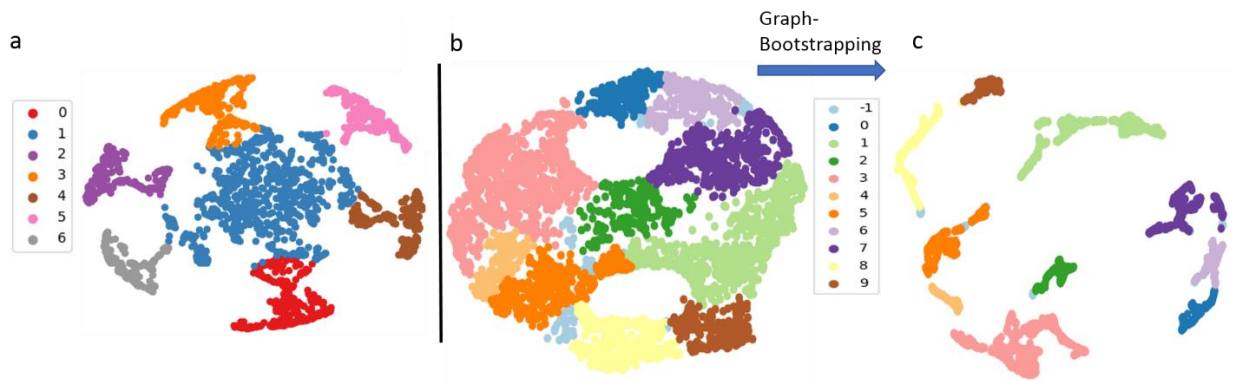

**Figure 2**: Manifold bootstrapping and clustering. (a) Spectral clustering results on UMAP manifold derived from the synthetic dataset. (b) UMAP manifold and (c) Bootstrapped UMAP manifold derived from the experimental dataset. UMAP and bootstrapped UMAP manifolds are colored by the same set of clustering labels where HDBSCAN clustering was performed on the bootstrapped manifold. Bootstrapped manifold in (c) is derived from the reconstructed graph based on the (b) UMAP manifold.

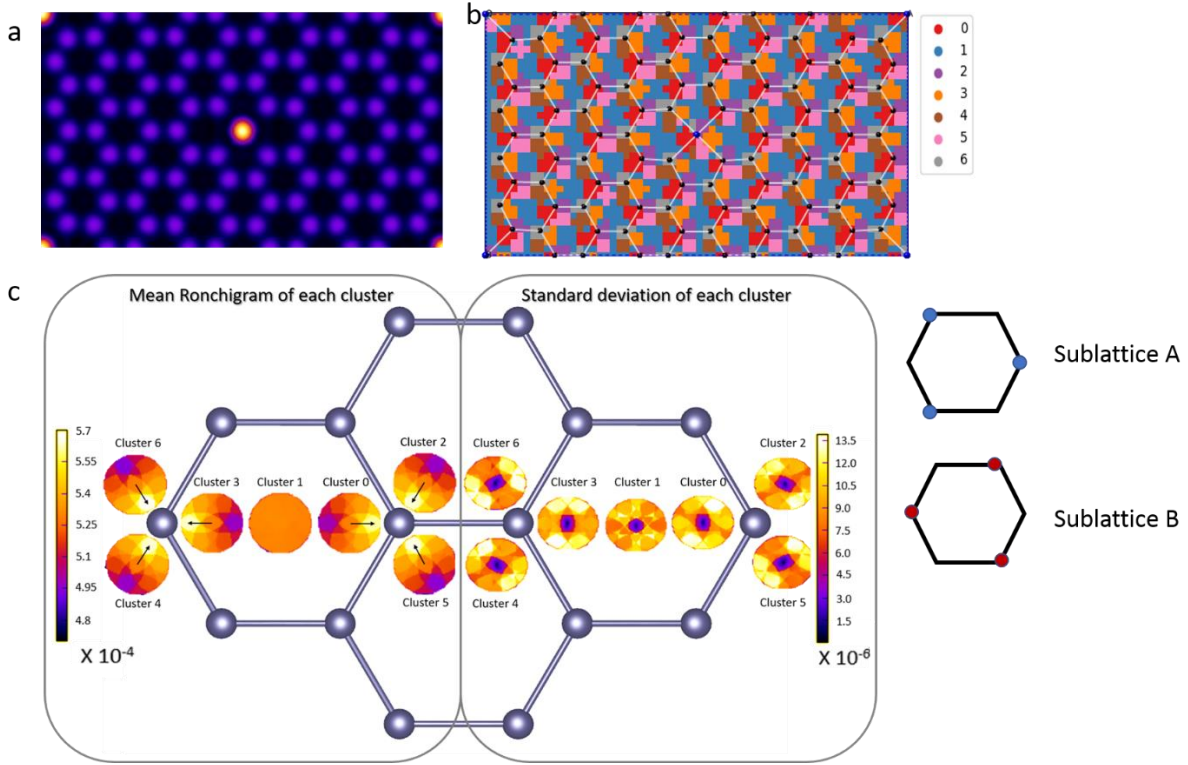

**Figure 3**: Synthetic data analysis. (a) HAADF image of the synthetic dataset. (b) Ground-truth atom positions overlaid on real-space distributions of cluster labels in Figure 2a. Black atoms are carbon, and blue ones are the Si dopants. (c) The mean and standard deviation (Std) of Ronchigrams for each cluster and positions to the atom sites. Arrows indicate the simulated beam deflection observed near an atom. The Ronchigram intensity moves toward the atom.

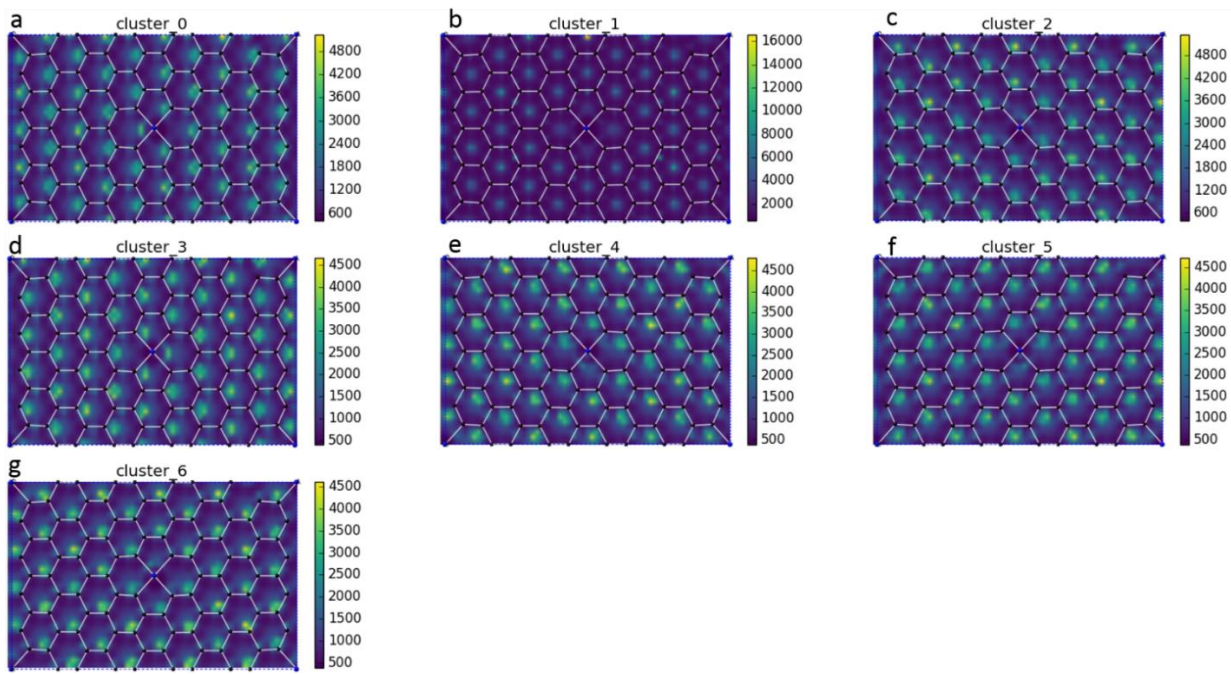

**Figure 4**: Similarity loadings of clusters from the synthetic dataset. (a,c,f) Similarity loadings of clusters 0,2,5 over sublattice A. (b) Similarity loadings of cluster 1 located in the space between atoms. (d,e,g) Similarity loadings of clusters 3,4,6 over sublattice B.

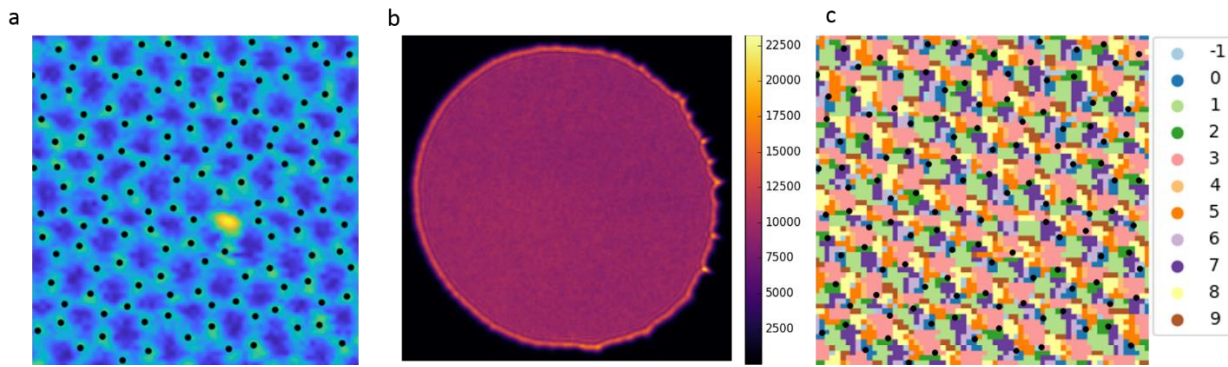

**Figure 5**: Experimental data analysis. (a) The HAADF image overlaid with atom positions. (b) An as-acquired experimental Ronchigram image, corresponding to the top-left pixel position in (a). (c) Real-space distributions of cluster labels in Figure 2c.

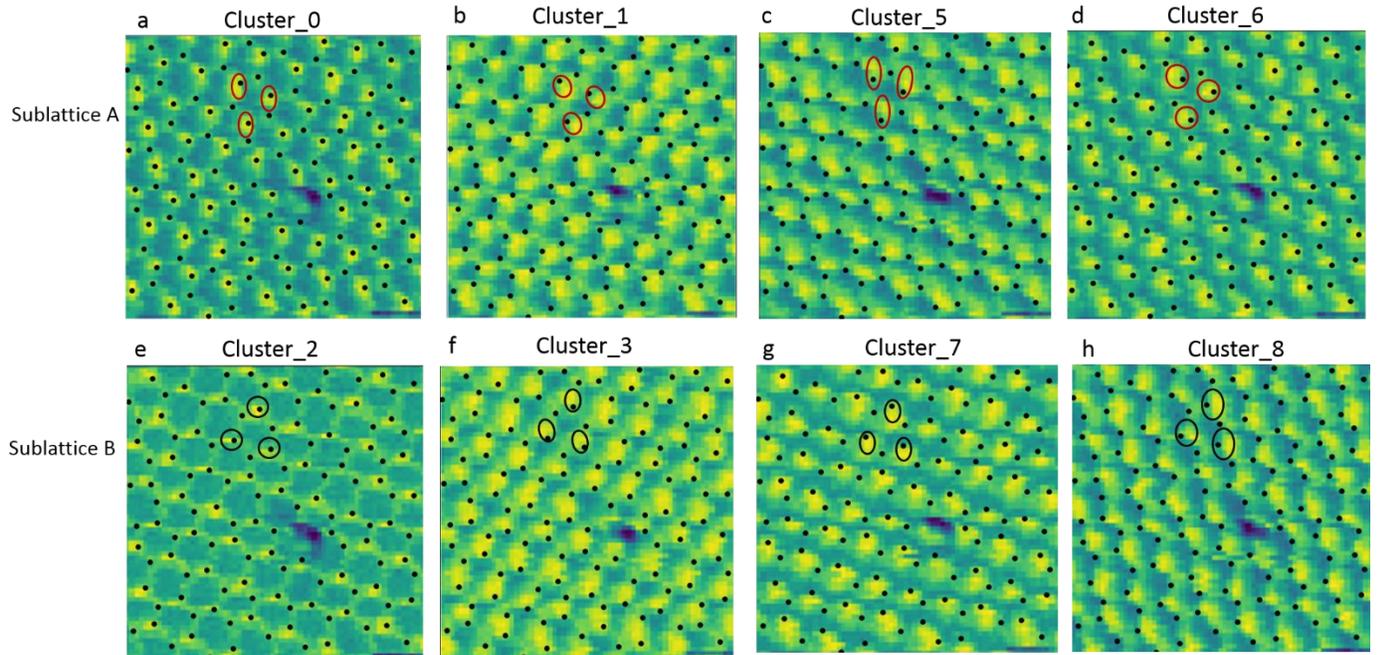

**Figure 6**: Similarity loadings of clusters from the experimental dataset. (a-d) Similarity loadings of clusters over sublattice A. (e-h) Similarity loadings of clusters over sublattice B. Here we overlay the circle markers to illustrate the relative positions between atom sites and bright blobs.

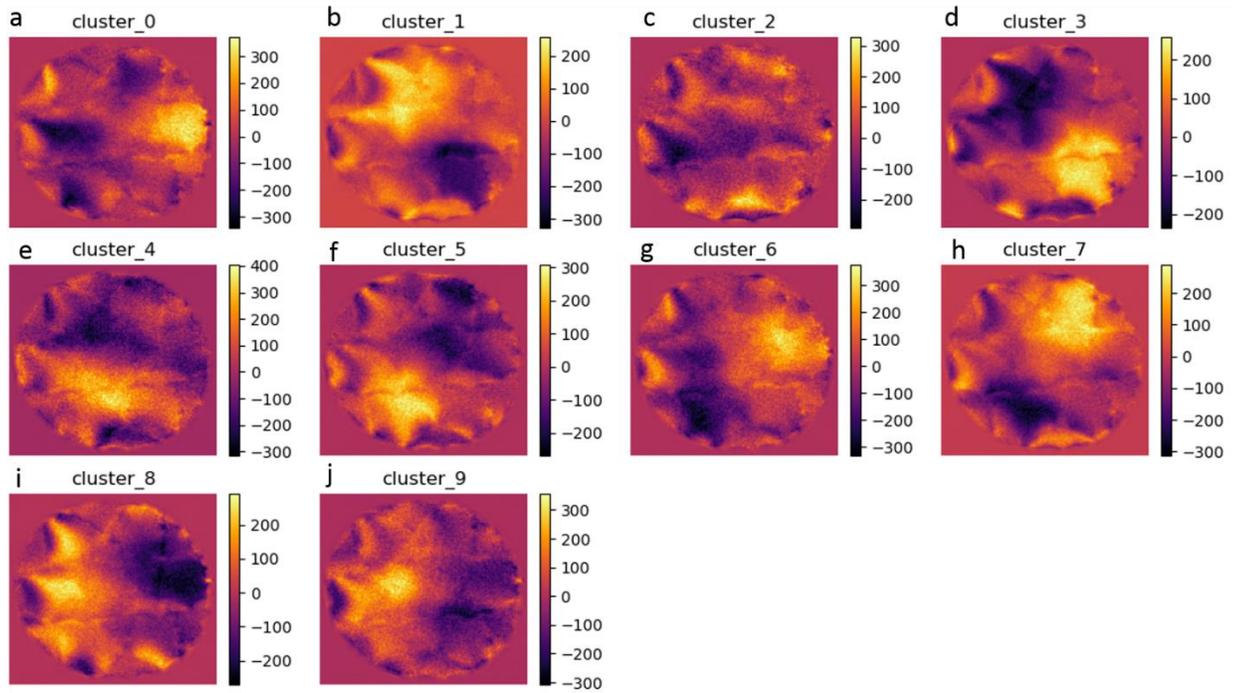

**Figure 7**: Deflection patterns in cluster Ronchigrams from the experimental dataset. (a-j) Deflection patterns in mean Ronchigrams for clusters 0-9. Here we subtracted the mean of all Ronchigrams from mean Ronchigram of each cluster.

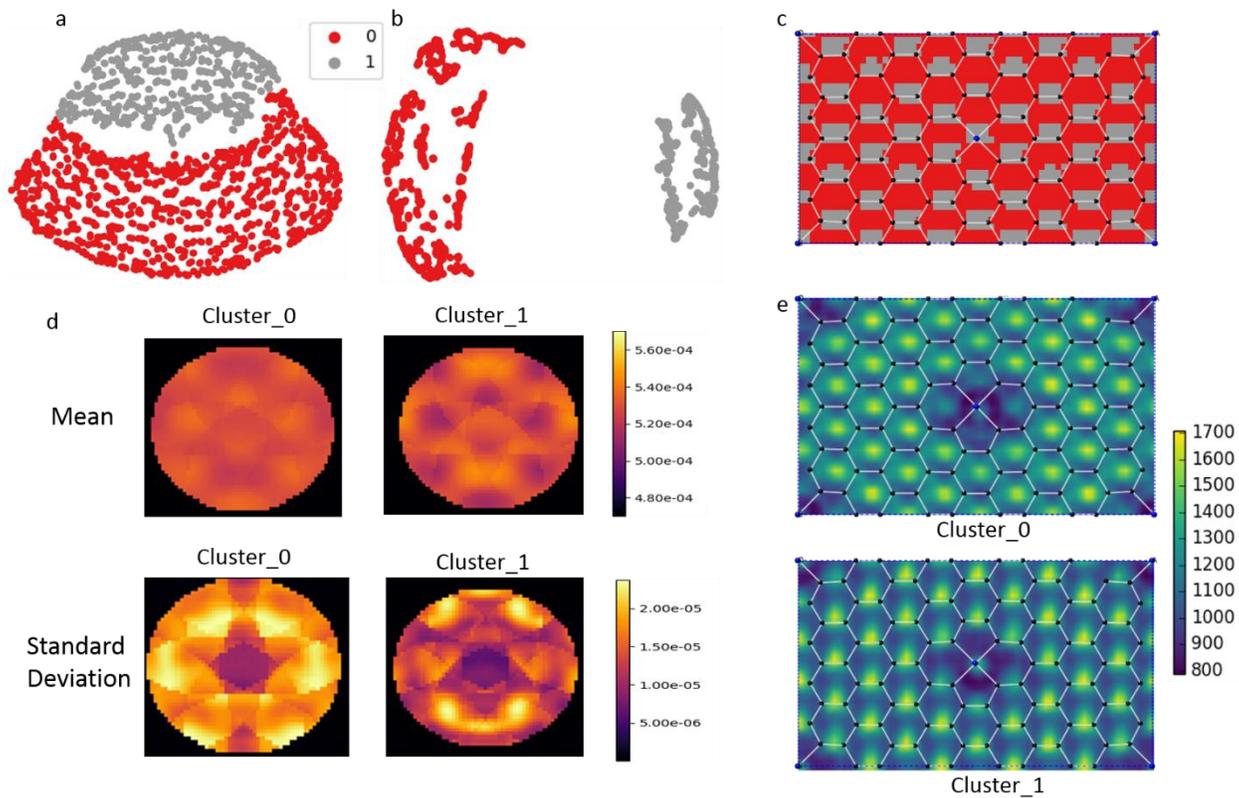

**Figure 8:** Synthetic data analysis with defocus setting. (a) UMAP and (b) bootstrapped UMAP manifolds. (c) Real-space distributions of cluster labels. (d) The mean and standard deviation (Std) of Ronchigram for each cluster. (e) Similarity loadings of the clusters.

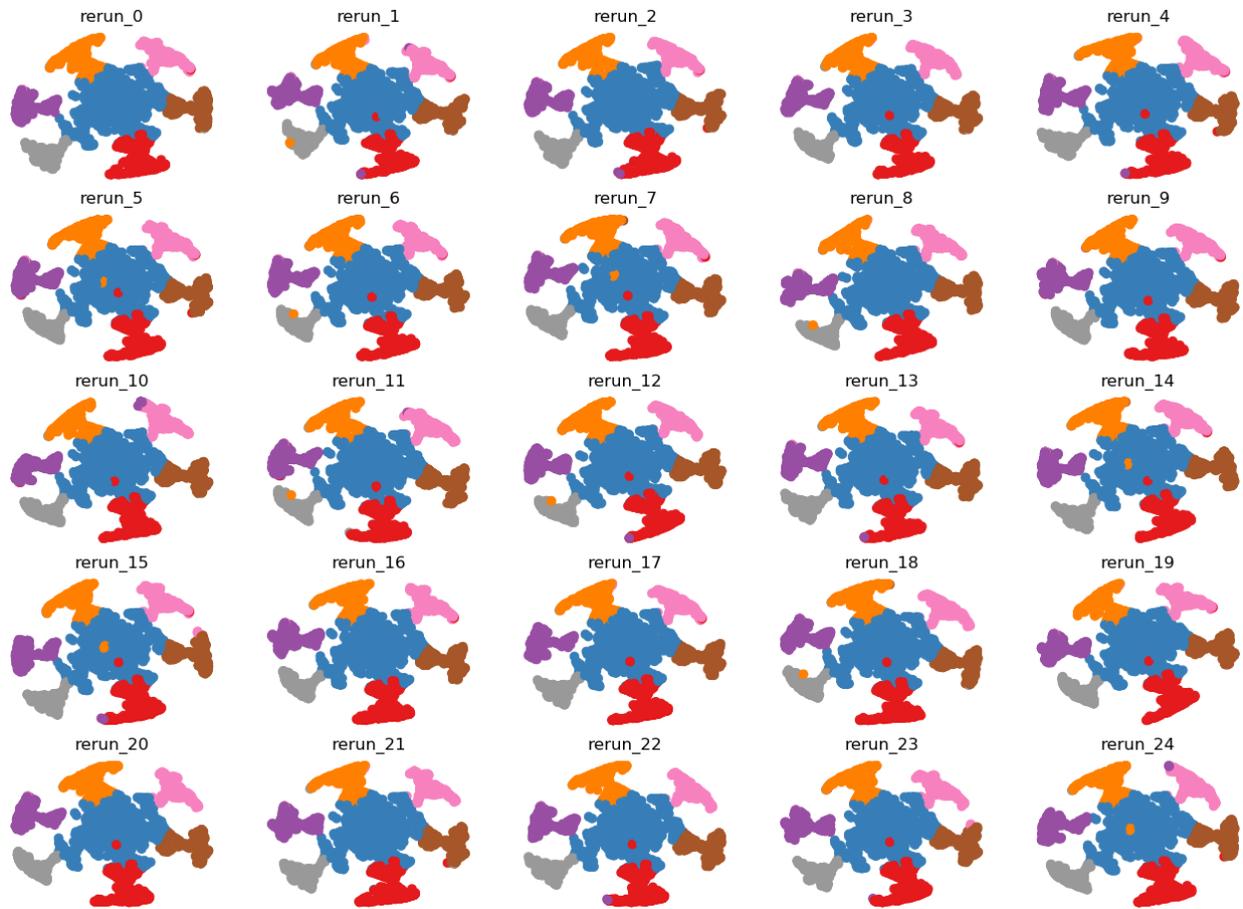

Supplementary Figure 1: Manifold patterns of 25 reruns of UMAP on the synthetic dataset. Here we overlay the same set of cluster labels in Figure 2a.

Supplementary Figure 2: Trend of total number of HDBSCAN clusters on bootstrapped UMAP manifold on the experimental dataset, fitted by the exponential decay in the form of $P(k) = Ce^{-(k-20)/\tau} + b$.

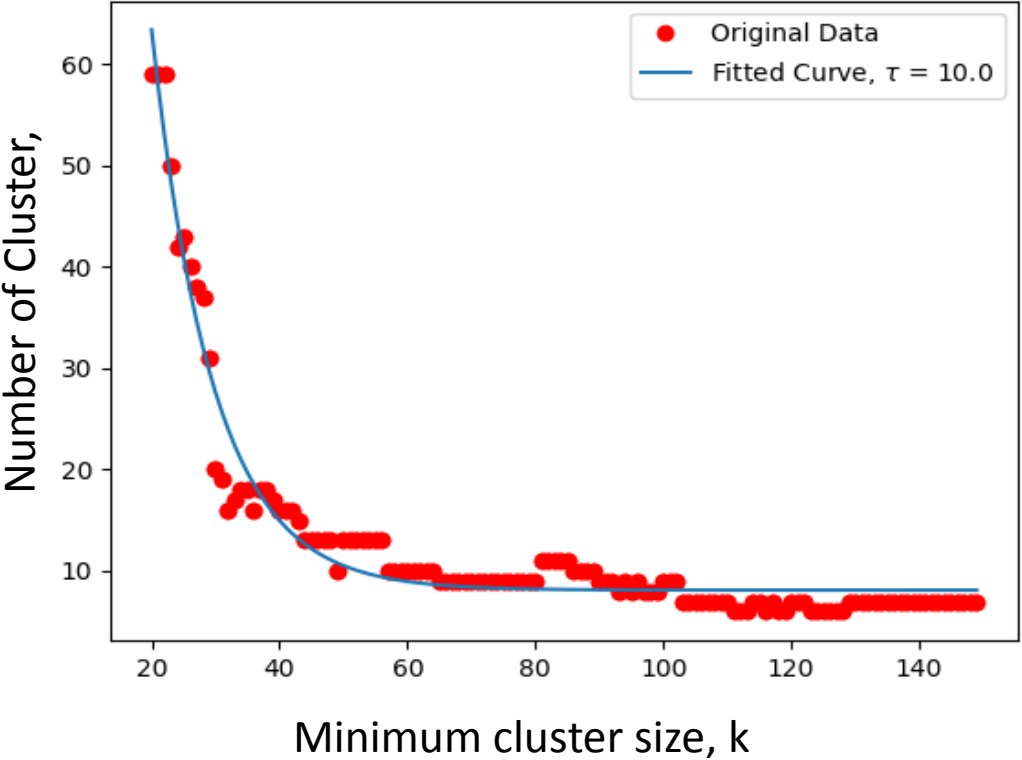

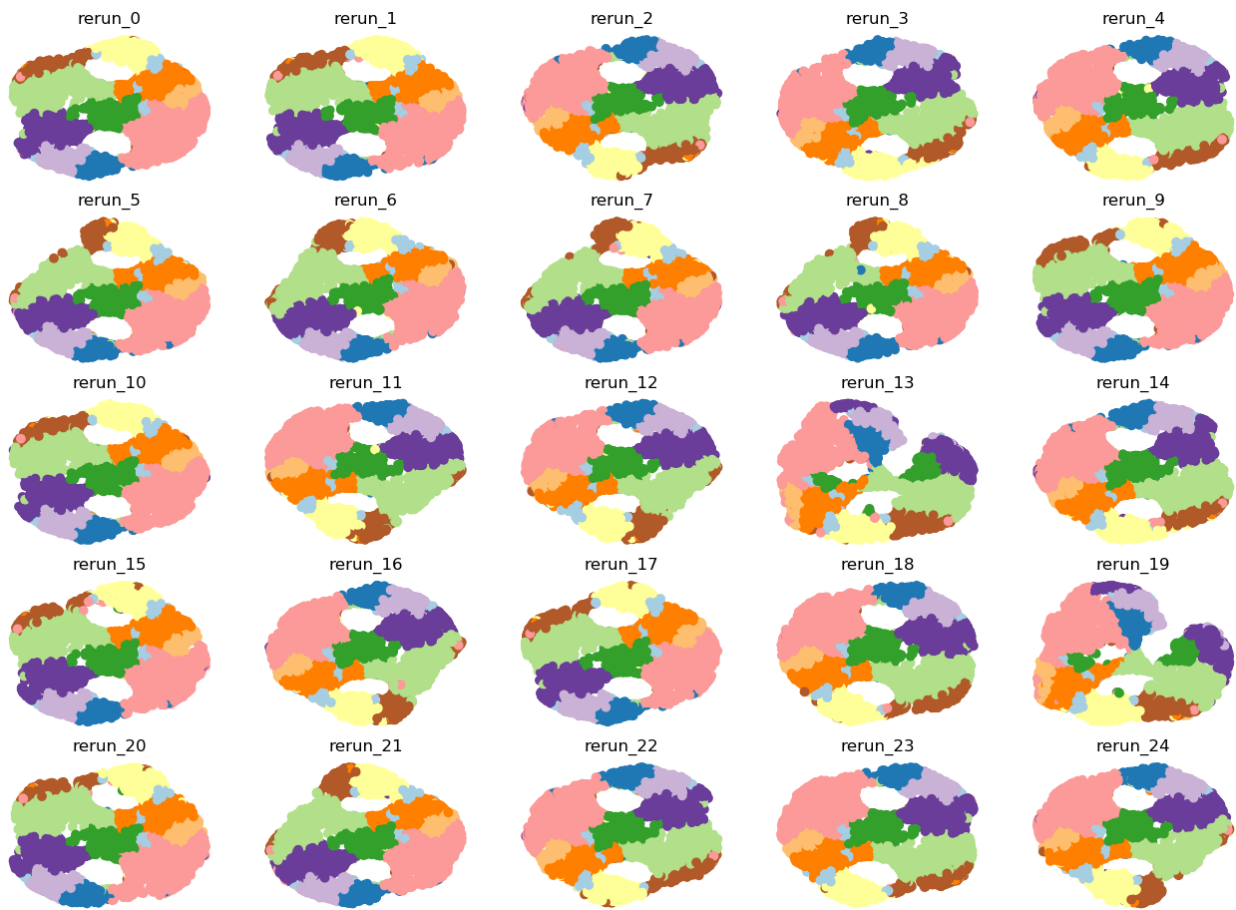

Supplementary Figure 3: Manifold patterns of 25 reruns of UMAP on the experimental dataset. Here we overlay the same set of cluster labels in Figure 2b,c

Supplementary Figure 4: Similarity loadings of clusters from the experimental data, whose locations are not indicative to either sublattice. (a,b) Clusters 4,9.

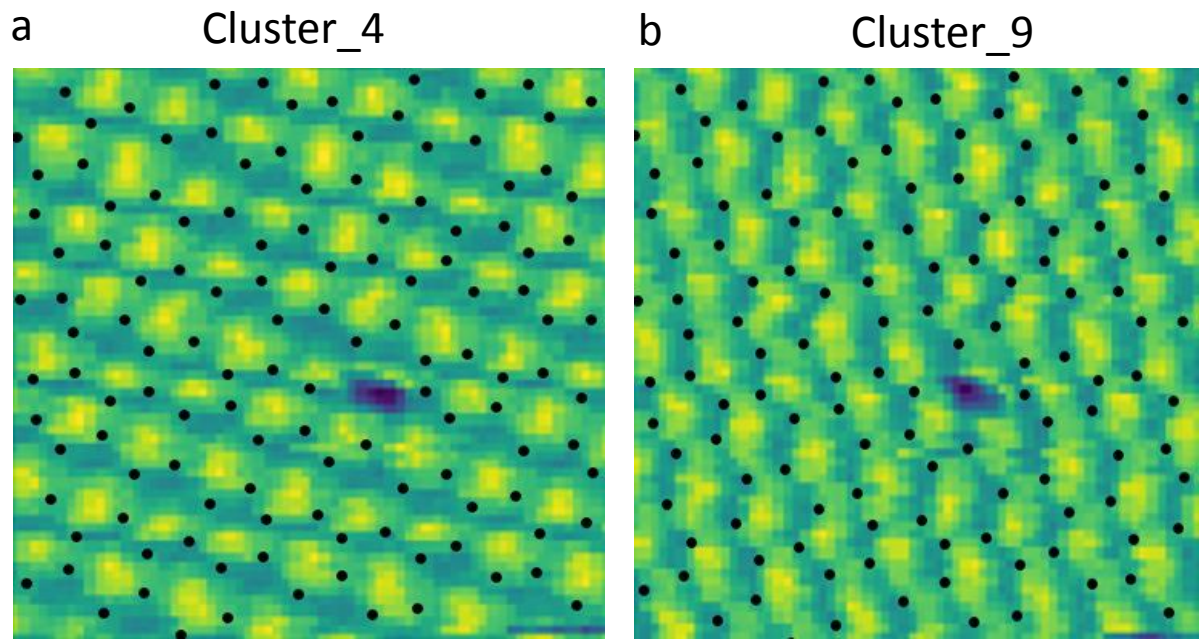

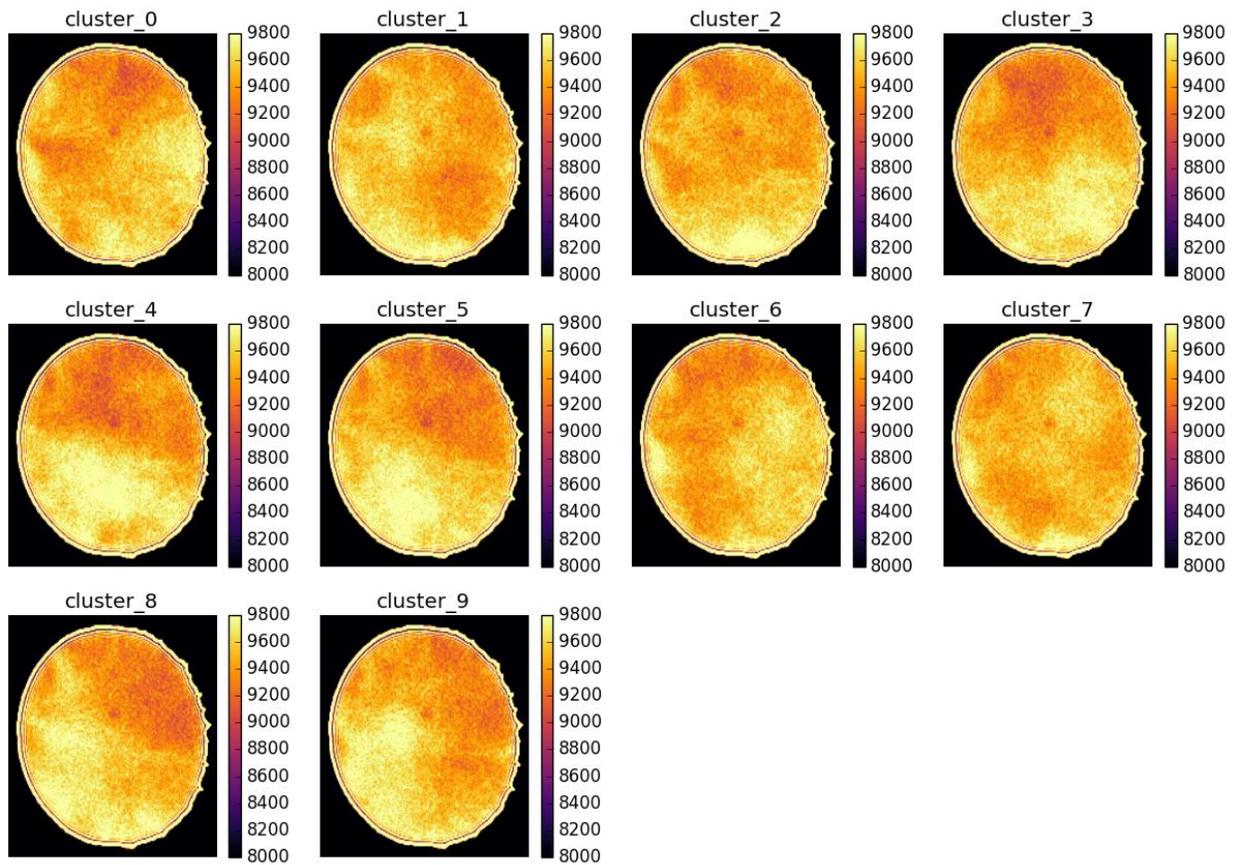

Supplementary Figure 5: Deflection patterns in the mean Ronchigram of each cluster from the experimental dataset. Here we manually tuned the range of colorbar to be [8000, 9800].

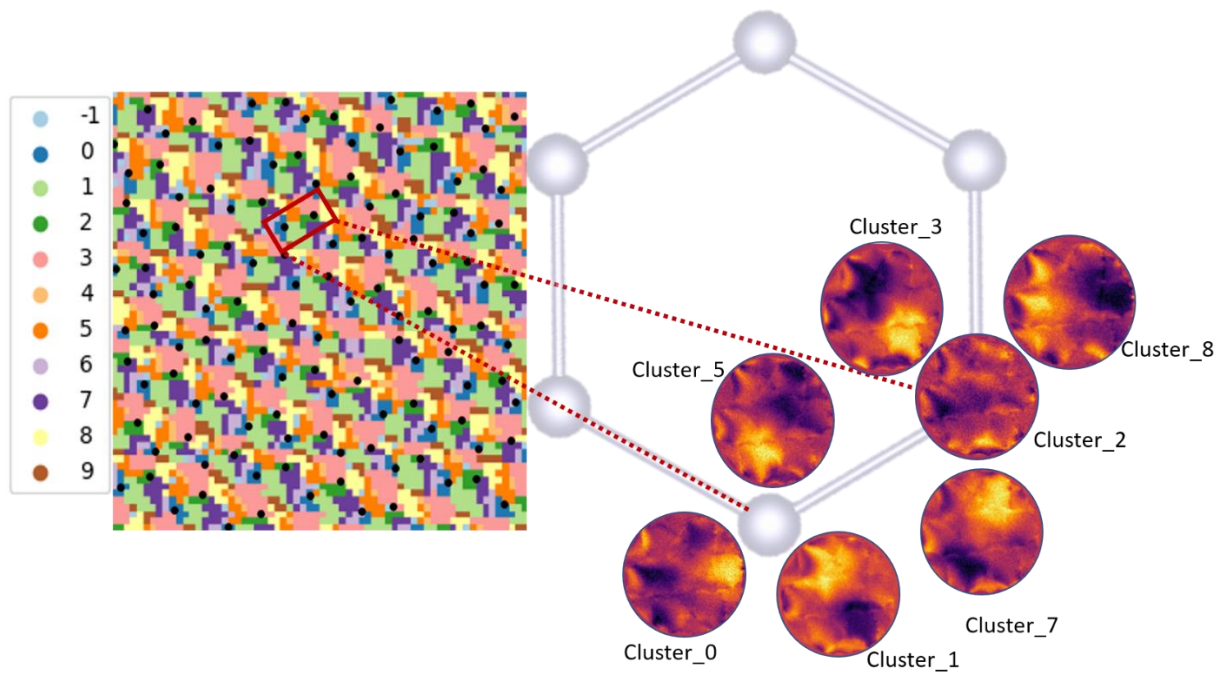

Supplementary Figure 6: Illustration of deflection patterns in experimental Ronchigrams and associated positions to the atom sites.